\documentclass[
reprint,
amsmath,
amssymb,
aps,
prl,
longbibliography,
floatfix
]{revtex4-2}

\usepackage{graphicx}
\usepackage{color}
\usepackage[english]{babel}
\usepackage[T1]{fontenc}
\usepackage[utf8]{inputenc}
\usepackage{dcolumn}
\usepackage{bm}
\usepackage{hyperref}
\usepackage[capitalise]{cleveref}
\usepackage[mathlines]{lineno}
\usepackage{multirow}
\usepackage{braket}

\def\ket#1{| #1 \rangle}
\def\bra#1{\langle #1|}

\newcommand{\ci}{\mathrm{i}}

\usepackage{MnSymbol}
\usepackage{physics}

\begin{document}

\preprint{APS/123-QED}

\title{Mollow-like triplets in ultrafast resonant absorption}

\author{Axel Stenquist$^1$\,
}
\author{Felipe Zapata$^2$*\,
}
\author{Edvin Olofsson$^1$\,
}
\author{Yijie Liao$^3$\,
}
\author{Elna Svegborn$^1$\,
}
\author{Jakob Nicolai Bruhnke$^1$
}
\author{Claudio Verdozzi$^1$\,
}
\author{Jan Marcus Dahlström$^{1\text{\dagger}}$\,
}

\affiliation{$^1$Department of Physics, Lund University, 22100 Lund, Sweden. \\
$^2$Departamento de Química, Universidad Autónoma de Madrid, 28049 Madrid, Spain. \\
$^3$School of Physics and Wuhan National Laboratory for Optoelectronics,
Huazhong University of Science and Technology, Wuhan, 430074, China.}

\begin{abstract}
\noindent 
We show that resonant absorption of smooth laser fields can yield Mollow-like triplet patterns. General conditions for such triplets are derived and illustrated with a super-Gaussian pulse sequence. Gaussian pulses can not exhibit triplets, super-Gaussian pulses can form triplets depending on the pulse area and flat-top pulses can produce absorption triplets after one Rabi cycle. Our results are compared side-by-side with resonance fluorescence to emphasize similarities and differences between these unlike observables. In the high-intensity limit, we show that the central absorption peak is asymmetric, which we attribute to non-linear photoionization, beyond two-level atomic physics.
\end{abstract}

\maketitle

\begin{table}[b!]
\begin{flushleft}
  * felipe.zapata@uam.es
  
  \noindent
  $^\text{\dagger}$ marcus.dahlstrom@matfys.lth.se
\end{flushleft}
\end{table}

Mollow triplet formation is a well-known phenomenon in resonance fluorescence from atoms \cite{mollow_power_1969}, with a characteristic three-peaked spectrum. The structure can be understood from the dressed-state picture with  Quantum Electrodynamics (QED) for continuous wave lasers \cite{cohen-tannoudji_modification_1977,kimble_theory_1976},
where adjacent peaks are separated by the Rabi frequency. While Mollow triplets in resonance fluorescence have been observed in many quantum systems, such as highly-charged ions \cite{cavaletto_resonance_2012}, cold atoms \cite{ortiz-gutierrez_mollow_2019}, quantum dots \cite{ulhaq_cascaded_2012,ulhaq_detuning-dependent_2013} and in optically confined atoms \cite{ng_observation_2022}, the phenomenon has been predicted to be strongly modified for smooth laser pulses of finite duration due to interference of photons from different times \cite{rzazewski_resonance_1984,florjaczyk_resonance_1985,lewenstein_theory_1986,lewenstein_probing_1993,cavaletto_resonance_2012,moelbjerg_resonance_2012,vinas_bostrom_time-resolved_2020}. Experimental evidence for such dynamically dressed states by resonance fluorescence was recently found in quantum dots \cite{boos_signatures_2024,liu_dynamic_2024}. 

During the last decade, transient absorption spectroscopy has been used to measure coherent electronic dynamics in atoms and molecules with unprecedented resolution down to the attosecond timescale \cite{goulielmakis_real-time_2010,wang_attosecond_2010,wirth_synthesized_2011,ott_lorentz_2013,ott_reconstruction_2014,beck_probing_2015,birk_attosecond_2020,marroux_attosecond_2020,rupprecht_laser_2022,wu_theory_2016,stenquist_gauge-invariant_2023,baggesen_theory_2012,pabst_theory_2012}. 
However, Mollow-like absorption triplets have so far not been reported. This may be due to inadequate laser conditions that could prevent such effects from manifesting in absorption driven by strong laser fields, or due to some unknown physical restrictions. The rapid development of seeded extreme ultraviolet free electron lasers (XUV-FEL), such as FERMI \cite{allaria_highly_2012}, opens up for such studies, as evidenced by the recent observation of dressed-states between helium atoms and light at XUV wavelengths \cite{nandi_observation_2022,nandi_generation_2024}. Whether Mollow-like triplets in absorption spectra can be generated from seeded XUV-FEL pulses is thus a timely question that requires theoretical attention.

\begin{figure*}[t!]
    \centering
    \includegraphics[width = 1\textwidth]{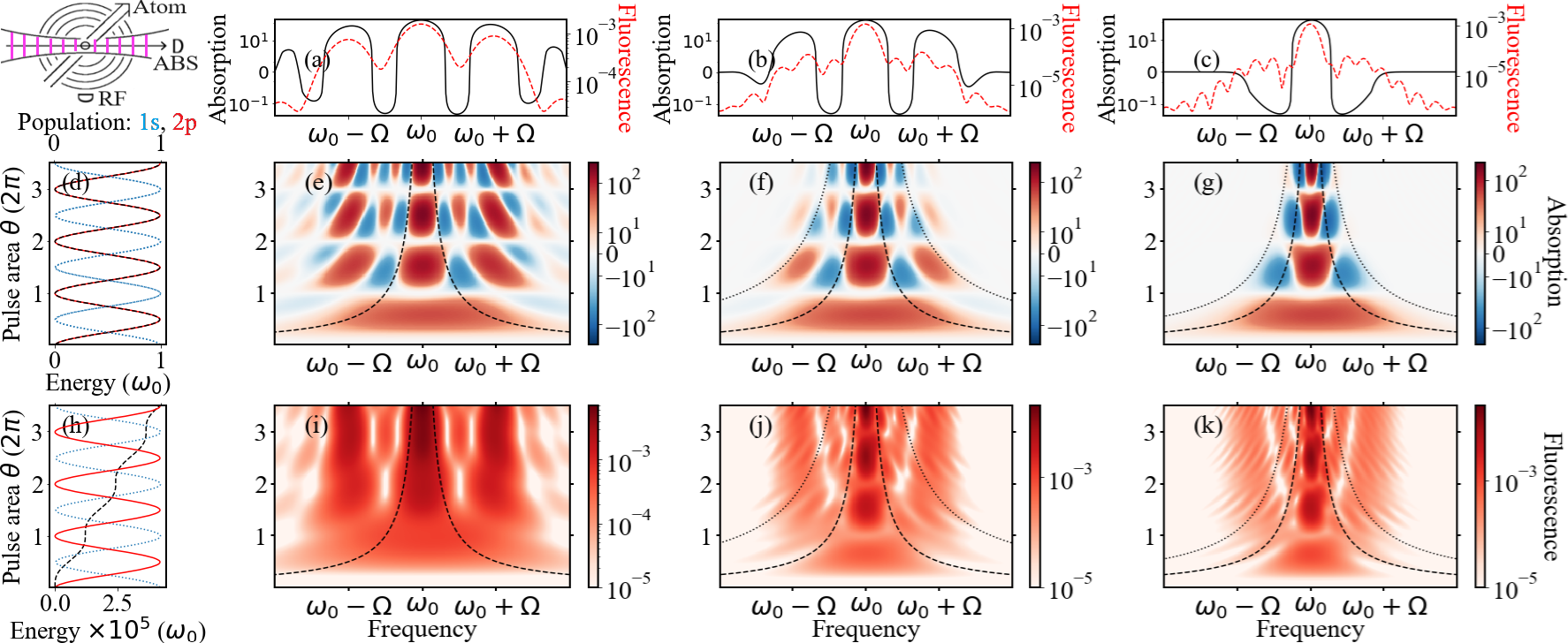}
\caption{Absorption (ABS) and resonant fluorescence (RF) spectra are presented for a fixed intensity of the laser, $I = 10^{12}$ W/cm$^2$, yielding the Rabi period $T_{Rabi} = 38$ fs. In the top left, an illustration of ABS and RF measurements is presented. \textbf{First row:} Line outs present the ABS (black) and RF (red dashed) for 3/2 Rabi cycles ($3\pi$ pulse area) in (a), (b) and (c), for a flat-top, super-Gaussian and Gaussian envelope, respectively. \textbf{Second row:} ABS, resolved over pulse area, (d) showing the number of absorbed photons (black dashed) and the population of the excited (red) and ground state (blue dotted) for a flat top pulse. Panels (e), (f) and (g) show the frequency-resolved ABS for the flat top, super-Gaussian and Gaussian pulse shapes, respectively, in symmetric logarithmic scale. ABS is computed with a filter function equal to the pulse envelope. \textbf{Third row:} RF, resolved over pulse area, calculated by a model based on QED \cite{rzazewski_resonance_1984,florjaczyk_resonance_1985}, and scaled using the classically emitted radiation, see supplemental material \cite{Supplemental_Material}. The heatmaps are superposed with dashed lines that describe the energy width of the central peak of the field:  $2\pi/\tau$, and with dotted lines that describe a larger width of the pulse, $2\pi/\tau_n'$, see \cref{fig:Pulse shapes} and main text. }
\label{fig:MT_time}
\end{figure*}

In this letter, we employ strong-field transient absorption theory \cite{wu_theory_2016,stenquist_gauge-invariant_2023} for resonant XUV-FEL pulses in hydrogen atoms to elucidate the conditions for Mollow-like absorption triplets, resolved over the time, intensity and the shape of the laser pulse. Our results for absorption are compared with QED calculations for resonance fluorescence  \cite{rzazewski_resonance_1984,florjaczyk_resonance_1985}, showing both remarkable similarities and stark differences between these two unlike observables (see insert in Fig.~\ref{fig:MT_time}). In contrast to fluorescence, we show that the triplet structure in absorption is only formed when two seemingly opposing conditions can be simultaneously fulfilled. To study the dynamic build-up of Mollow-like absorption features, we consider the prototypical $1s-2p$ transition in hydrogen, which has a resonant energy of $\omega=10.2$\,eV, and a transition dipole element $z_0=0.745$. Atomic units are used throughout this text, $e=\hbar=m=4\pi\epsilon_0=1$, unless otherwise stated.

\textit{\textbf{Theory:}}
Absorption of light is computed semi-classically using the fundamental energy conservation condition between the atom and the driving laser field, ${\cal E}={\cal E}_\mathrm{A}(t)+{\cal E}_\mathrm{L}(t)$, following Wu et al. \cite{wu_theory_2016}. The time-dependent energy gain of the atom, $\Delta {\cal E}_\mathrm{A}(t) = - \Delta {\cal E}_\mathrm{L}(t)$, can be formally associated with Yang's energy operator \cite{yang_gauge_1976}, to write the theory in a gauge-invariant time-dependent form \cite{stenquist_gauge-invariant_2023}.  
The exact electron dynamics of a hydrogen atom is computed, within the dipole approximation, by solving the time-dependent Schrödinger equation,  
\begin{equation} \label{Eq: TDSE}
    \ci \frac{d}{dt} \ket{\Psi(t)} = [H_0 + V(t)]\ket{\Psi(t)},
\end{equation}
where $H_0=\mathbf{p}^2/2-1/r$ is the hydrogen Hamiltonian and $V(t)=zE(t)$ is the semi-classical interaction term. 
The electric field is linearly polarized along the z-direction and is defined by $E(t)=-\dot A$, with $A(t)=A_0\Lambda_n(t) \sin(\omega_0 t)$, where $A_0$ and $\omega_0$ are the amplitude and central frequency, respectively. The pulse envelope is defined using the super-Gaussian sequence \cite{florjaczyk_resonance_1985} as  
\begin{equation}\label{Eq: evelope}
    \Lambda_n(t) = \exp \left[ -\frac{\text{ln}(2)}{2} \left(\frac{2t}{\tau}\right)^{2n}\right],
\end{equation}
where $\tau$ is the pulse duration and $n=1,2,...$ the index of the sequence. Here $n=1$ corresponds to the commonly used Gaussian pulse, while the limit $n \rightarrow \infty$ corresponds to gradual transformation to a flat top pulse. 

The time-dependent velocity expectation value is computed as $\mathbf{v}(t) = \bra{\Psi(t)}\mathbf{v} \ket{\Psi(t)}$. 
In the frequency domain, the energy gain is resolved as \cite{stenquist_gauge-invariant_2023}
\begin{equation}
\begin{split}
\label{Abs_w}
    \Delta \tilde {\cal E}_\mathrm{A}(\omega)
    &= -2\;\mathrm{Re}[\tilde v_z(\omega)\tilde{E}^{*}(\omega)],
\end{split}
\end{equation}
where $\omega$ is the angular frequency, and $\tilde v_z(\omega)$ and $\tilde{E}(\omega)$ are the Fourier transforms of the velocity expectation value (along $z$) and the electric field, respectively.

In order to interpret the atomic dynamics we make use of the well-known two-level physics \cite{allen_optical_1987}. The Rabi flopping of a two-level atom follows the pulse area, $\theta=\int \Omega(t) dt$, where $\Omega(t) \approx \omega_0 A_0 \Lambda_n(t)z_0$ is the time-dependent Rabi frequency. We have verified that the two-level model is in excellent agreement with our exact results at $10^{12}$ W/cm$^2$, and below. Therefore, we present our results in terms of completed Rabi cycles: $\theta/2\pi$. Further details are given in the supplemental material \cite{Supplemental_Material} (see also references \cite{cohentannoudji_atomphoton_1998,olofsson_photoelectron_2023,paulisch_beyond_2014,faisal_theory_1987} therein). 

\textit{\textbf{Results:}} The pulse area of $3\pi$ is of particular interest. Physically, this corresponds to the atom being excited, de-excited, and re-excited by the laser field with net absorption of one photon ($\omega_0$) at the end of the laser pulse. The corresponding results for absorption and resonant fluorescence with a flat-top ($n\rightarrow\infty$), super-Gaussian ($n=2$) and Gaussian ($n=1$) envelope are presented in the first row of \cref{fig:MT_time}. In the following we will present results for each envelope in turn: (a)-(c).

\textit{Flat-top.--} Three strong absorption peaks (black) are observed in remarkable agreement with the corresponding resonance fluorescence result (red dashed), see \cref{fig:MT_time}~(a). 
Resonance fluorescence is here a non-negative quantity, because it only concerns emission of light, see schematic in \cref{fig:MT_time}. In contrast, absorption can take positive and negative values due to interference between the incident laser pulse and the stimulated emission,  in agreement with the pioneering work by Mollow for two-level atoms relaxed in monochromatic light subject to a perturbative auxiliary field \cite{mollow_stimulated_1972}. In our time-resolved non-perturbative approach, shown in (d), the energy absorbed by the atom modulates periodically by exchange of one photon (black dashed), following closely the excited state population (red). In (e) we show the frequency-resolved absorption from a flat-top pulse. Below one Rabi period, a single broad absorption peak is observed, which vanishes completely for a pulse area of exactly one Rabi period because the atom is driven back to the ground state. For pulse areas greater than one, we first observe a Mollow-like absorption triplet at 1.5 Rabi cycles. At 2.5 Rabi cycles, a new structure is formed with an additional weaker absorption peak in between the sidebands and the main peak. At 3.5 Rabi cycles, there are two additional absorption peaks on either side of the main absorption peak.

\begin{figure}[t!]
    \centering
    \includegraphics[width = 0.49\textwidth]{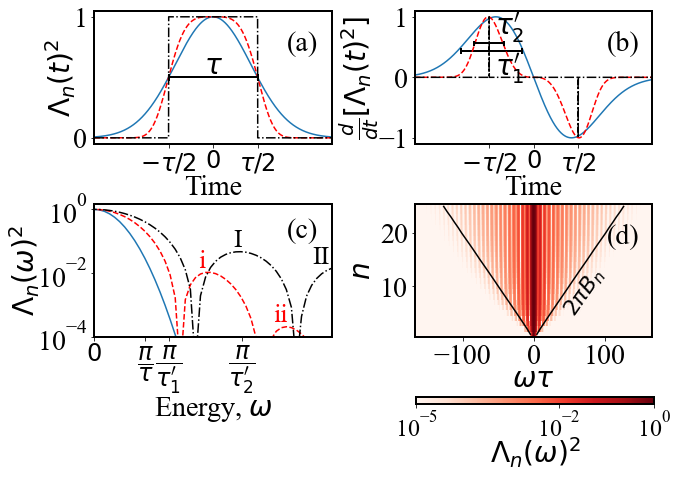}
\caption{Pulse structure characterized by pulse duration, $\tau$, and \textit{turn-on/off} duration, $\tau_n'$. Three pulses in the super-Gaussian sequence, given by \cref{Eq: evelope}, are shown: Gaussian ($n=1$) (blue), super-Gaussian ($n=2$) (red dashed) and flat top ($n \rightarrow \infty$) (black dash-dotted). The squared envelope (a) and its derivative (b) are presented with the corresponding characteristic times, $\tau>\tau_n'$. (c) Squared Fourier transformed envelope with energy half width of the central component, $\pi/\tau$, and a larger measure for the width, $\pi/ \tau_n'$. (d) Same as (c) resolved over super-Gaussian index $n$, where black lines are $\pm\pi\tau/\tau'_n = \pm 2 \pi \mathcal{B}_n$.}
\label{fig:Pulse shapes}
\end{figure}

\textit{Super-Gaussian.--} Three Mollow-like absorption peaks (black) are observed at 1.5 Rabi cycles in \cref{fig:MT_time}~(b).
However, this traditional-looking Mollow structure is a special case that only occurs for the $3\pi$-area super-Gaussian pulse. This becomes clear when studying the frequency-resolved absorption, as a function of the pulse area, shown in \cref{fig:MT_time}~(f). Above 2 Rabi cycles, the structure deviates completely from the Mollow triplet, because the prominent sidebands, positioned at $\omega_0\pm\Omega_0$, are lacking or inferior to more narrow sidebands. 
Thus, at large pulse areas, the sidebands of super-Gaussian envelopes do not correspond to Mollow-like absorption structure.
\textit{Gaussian.--} At 1.5 Rabi cycle area the absorption spectrum shows only a single peak with two emission peaks (negative absorption) on either side as shown in \cref{fig:MT_time}~(c). 
In (g) we show that the frequency-resolved absorption over the pulse area maintains this shape, with the expected narrowing of the central peak due to the time-bandwidth product. Revivals of absorption are found at 2.5 and 3.5 Rabi cycles. All side peaks from the super-Gaussian case (f) have vanished in the Gaussian case (g). This implies that the Gaussian pulse can not support the $3\pi$-area Mollow triplet formation observed in the super-Gaussian case. Instead, the spectrum exhibits a central absorption peak and two emission peaks -- a distinct feature that is different from Mollow triplet formation.

\textit{Fluorescence.--} In contrast to absorption, energy lost by resonance fluorescence is very small and increases approximately linearly with the pulse area (for fixed intensity), as shown in (h). We find that the energy lost through fluorescence is negligible (five orders of magnitude smaller than the absorption), validating the fundamental assumption for energy conservation of the combined semi-classical atom and laser system, ${\cal E}$, \cite{wu_theory_2016,stenquist_gauge-invariant_2023}. 
Similarly, this also validates the omission of back action in QED from fluorescent photons \cite{rzazewski_resonance_1984,florjaczyk_resonance_1985}, thus circumventing the need for more complete QED treatment, c.f. Ref. \cite{lewenstein_theory_1986}.
The Mollow triplet in fluorescence forms with the expected strong central peak and two sidebands, as shown in (i). We note that the weak final state dependency \cite{moelbjerg_resonance_2012,boos_signatures_2024} is in strong contrast with our absorption results (e). In between the central peak and the sidebands, the Mollow structure develops additional peaks at 3 Rabi cycles (i), reminiscent of the additional absorption peaks observed at 2.5 Rabi cycles in (e). In the super-Gaussian (j) and Gaussian (k) cases the fluorescence shows a more complex structure with a central peak with broad wings on either side, which exhibits rich interference fringes  \cite{moelbjerg_resonance_2012,liu_dynamic_2024,boos_signatures_2024}.
In contrast to the fluorescence, associated to QED, the absorption is very different as it is not perturbative, but  follows instantaneously the population of the excited state on the femtosecond timescale, as observed in (d).  

\begin{figure*}[t!]
    \centering
    \includegraphics[width = 0.99\textwidth]{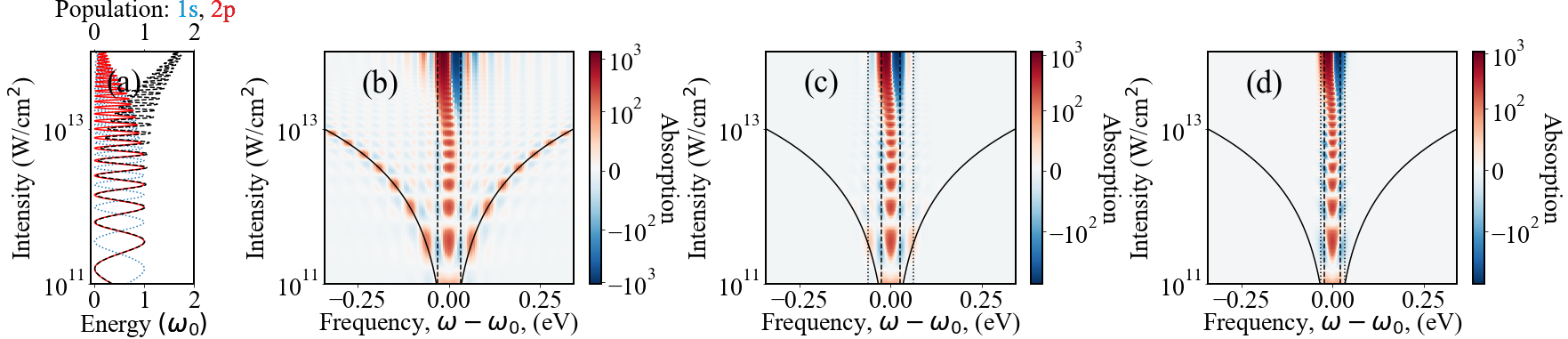}
\caption{Absorption spectra resolved over intensity, in the interval $I = 10^{11}-10^{14}$ W/cm$^2$ for the pulse duration $\tau = 97$ fs. The number of absorbed photons (black dashed) and the population of the excited (red) and ground state (blue dotted), for interaction with a flat top pulse, is shown in (a). We observe two photon absorption for high intensity, due to photoionization. Frequency-resolved absorption is shown in (b), (c) and (d) for a flat-top, super-Gaussian ($n=2$) and Gaussian pulse, respectively. Full black lines denote the energies $\pm \Omega$, i.e. the Mollow triplet separation, dashed vertical lines are the energy width of the field's central component $\pi/\tau$ and dotted lines are the full Fourier width of the pulse $\pi/\tau_n'$. }
\label{fig:MT_I}
\end{figure*}

\textit{\textbf{Discussion:}} 
The considered pulse envelopes, $\Lambda_n(t)$, are illustrated in \cref{fig:Pulse shapes}~(a) and (b) with their pulse duration, $\tau$, and {\it turn-on/off} duration, $\tau'_n$. To understand the envelope-dependence of the resonant absorption, presented in Fig.~\ref{fig:MT_time}~(e)-(g), we must appreciate these two different time-scales of the envelopes, $\tau$ and $\tau'_n$, which are associated with different widths in the frequency domain, $2\pi/\tau$ and $2\pi/\tau_n'$, as shown in \cref{fig:Pulse shapes}~(c) and (d). 
The central peak of the Gaussian is narrower in energy than the super-Gaussian and flat-top cases, and its half spectral width is well estimated as $\pi/\tau$ (at which point  its square has decreased by two orders of magnitude). The Gaussian also maintains its functional identity from time to energy, which explains why it does not exhibit any modulations in the energy domain beyond the primary peak in contrast to the other envelopes shown. 
The shorter turn-on/off duration of a super-Gaussian pulse leads to frequency modulations, labelled by $\mathrm{i},$ and $\mathrm{ii}$ in Fig.~\ref{fig:Pulse shapes}~(c). These additional peaks in the frequency domain are rather weak (its square decreases by two orders between adjacent maxima). For the flat-top pulse, the turn-on/off effect is much more significant, and it leads to several additional Fourier peaks on the same order of magnitude, labelled by $\mathrm{I}$ and $\mathrm{II}$. It is the presence of such modulations in the frequency domain that enables absorption features at large separations from the central frequency, by Eq.~(\ref{Abs_w}), see Fig.\ref{fig:Pulse shapes}~(d). 

We find that a pulse can support the Mollow-like absorption peaks under two conditions. i) the atom must undergo one Rabi cycle, $\theta > 2\pi $. In the frequency domain, this first condition implies that the bandwidth of the pulse is narrow enough to resolve the Rabi frequency. Combining the pulse area, $\theta = \Omega_0 \int \Lambda_n(t) dt$, where $\Omega_0 = 2\pi / T_{Rabi}$, with the flat top pulse area, $\int \Lambda_\infty(t)dt = \tau$, yields the condition for the lower bound of the pulse duration $\tau > {\cal A}_n T_{Rabi}$ where the dimensionless area constant is defined as ${\cal A}_n = \int \Lambda_\infty(t)dt / \int \Lambda_n(t) dt$.
ii) the pulse bandwidth must extend over the Rabi frequency, $\pi/\tau'_n > \Omega_0 \Leftrightarrow \tau < T_{Rabi}\, {\cal B}_n$, where we introduce a dimensionless constant, ${\cal B}_n = \tau/2\tau'_n\approx 0.8 n$, see Fig.\ref{fig:Pulse shapes}~(d). Thus, the Mollow-like structure can only develop under the general condition: 
\begin{equation}
\label{condition}
{\cal A}_n < \tau/T_\mathrm{Rabi} <  {\cal B}_n. 
\end{equation}
From this condition we find that Gaussian pulses can not support Mollow-like absorption triplets, since ${\cal A}_1= 1.50 > {\cal B}_1= 0.735$. For the super-Gaussian ($n=2$) case, however, the condition can be fulfilled, since ${\cal A}_2=1.18 < {\cal B}_2 = 1.45$, which allows for the first Mollow-like absorption triplet at $\theta/2\pi\approx 1.5$ in \cref{fig:MT_time}(f). For increasing super-Gaussian order, the condition is further relaxed. For the flat top case ($n\rightarrow\infty$), the turn-on time is infinitesimal, hence ${\cal A}_\infty=1< {\cal B}_n\rightarrow \infty$, which suggests that the modulations of the Fourier transform support Mollow-like absorption peaks at areas larger than one Rabi cycle, $\theta/2\pi>1$, in agreement with our observations in Fig.~\ref{fig:MT_time}~(e). Here multiple peaks are resolved on a colour scale that maps the first two orders of magnitude, each corresponding to the Fourier peaks $\mathrm{I}$, $\mathrm{II}$, etc, from Fig.~\ref{fig:Pulse shapes}~(c), see also (d).

\textit{The role of intensity.--} The resonant absorption for different pulse intensities is presented in \cref{fig:MT_I}. At low intensity, we see one photon absorption (black dashed line) in (a). As the intensity is increased, the stronger interaction with the laser leads to exchange energies beyond one photon, entering into the photoionization regime.  
Ionization yields decreased atomic 2-level populations as seen in the excited (ground) state population marked with a red (blue dotted) line. The frequency-resolved absorption spectra are shown in (b), (c) and (d) for interaction with a flat-top, super-Gaussian and Gaussian pulse, respectively. 
We observe that the flat-top case supports Mollow-like absorption triplets at all considered intensities. At very high intensity, however, the structure of the main peak changes into an asymmetric shape, with absorption below and emission above the resonant frequency, presenting a Fano-like profile. This asymmetric shape can be reproduced well by the damping of the amplitudes in an effective Hamiltonian two-level model, see supplemental material \cite{Supplemental_Material}. In the super-Gaussian case (c), only the first absorption triplet is formed ($3\pi$-area pulse), in agreement with Eq.~(\ref{condition}). Finally, in the Gaussian case (d), a single absorption peak is observed with emission structures on either side. Similar to the flat-top case (b), both the super-Gaussian (c) and Gaussian (d) cases exhibit a Fano-like asymmetric main absorption peak.

\textit{\textbf{Conclusions:}} In this letter, we have presented the conditions for Mollow-like triplets in resonant absorption of smooth laser fields. It was found that Gaussian pulses can not exhibit the triplet phenomenon, while super-Gaussian and flat-top pulses can produce triplets in absorption. However, the triplet structure of super-Gaussian pulses depends on the pulse area, which is a phenomenon beyond the usual Mollow picture. Side-by-side comparison with resonance fluorescence was made to emphasize the stark differences in the unlike physical observables. Our work, which lies on the boundary between semi-classical physics and QED, motivates experiments to study absorption of intense seeded XUV-FEL pulses, opening the possibility of probing ultrafast dynamics in  single- and multiphoton regimes in complex targets, such as molecules  \cite{Alicia2006,Pan2023,Chen2024} and highly-charged ions \cite{Postavaru2018}. Finally, a natural extension of the present work is developing a fully quantum treatment of the absorption theory, as an addition to the active fields of strong-field QED \cite{stammer_entanglement_2024,stammer_quantum_2023}.

\section{Acknowledgements} 

JMD acknowledges support from the Olle Engkvist Foundation: 194-0734 and the Knut and Alice Wallenberg Foundation: 2019.0154. CV acknowledges support from the Swedish Research Council (grant number 2022-04486) FZ acknowledges support from the Ministerio de Universidades, el Plan de Recuperación, Transformación y Resilencia y la Universidad Autónoma de Madrid: CA1/RSUE/2021-00352 and the MSCA H2020 programme: 101034324.

\bibliography{Mollow_ref}

\end{document}


\preprint{APS/123-QED}

\title{Supplemental Materials: \\
Mollow-like triplets in ultra-fast resonant absorption}
\author{Axel Stenquist$^1$\,
}
\author{Felipe Zapata$^2$*\,
}
\author{Edvin Olofsson$^1$\,
}
\author{Yijie Liao$^3$\,
}
\author{Elna Sveborn$^1$\,
}
\author{Jakob Nicolai Bruhnke$^1$
}
\author{Claudio Verdozzi$^1$\,
}
\author{Jan Marcus Dahlström$^{1\text{\dagger}}$\,
}
\affiliation{$^1$Department of Physics, Lund University, 22100 Lund, Sweden. \\
$^2$Departamento de Química, Universidad Autónoma de Madrid, 28049 Madrid, Spain. \\
$^3$School of Physics and Wuhan National Laboratory for Optoelectronics, Huazhong University of Science and Technology, Wuhan, 430074, China.}
  
\maketitle

\begin{table}[b!]
\begin{flushleft}
  * felipe.zapata@uam.es \\
  $^\text{\dagger}$ marcus.dahlstrom@matfys.lth.se
\end{flushleft}
\end{table}

\section{Transient Absorption Theory}\label{appendix}

In this work, we investigate the energy exchange between a hydrogen atom and a finite XUV-FEL pulse.  Atomic units are used throughout this text, $e=\hbar=m=4\pi\epsilon_0=1$, unless otherwise stated. 

\begin{figure}[t!]
    \includegraphics[width = 0.45\textwidth]{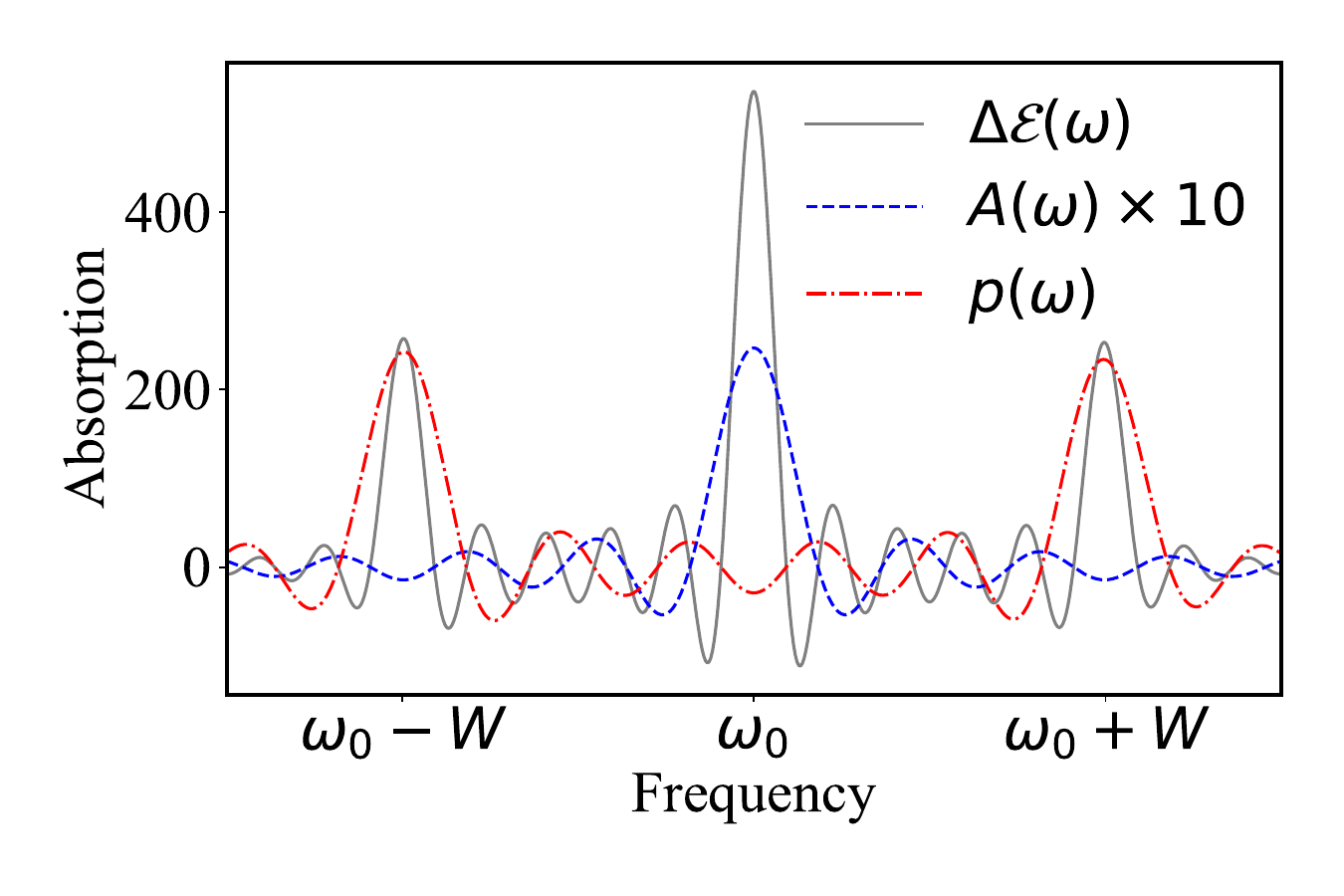}
\caption{Components of the Mollow-like absorption. The Fourier-transformed field potential (real part) and momentum expectation values (imaginary part) are presented along with the frequency-resolved absorption. Simulated using a flat top pulse with area $11\pi$ and CEP $\phi = \pi / 2$.}
\label{fig:Absorption component}
\end{figure}

The evolution of the electronic state, $\ket{\Psi(t)}$, is computed by solving the time-dependent Schrödinger equation (TDSE), 
\begin{equation} \label{Eq: TDSE}
    \ci \frac{d}{dt} \ket{\Psi(t)} = (H_0 + V(t))\ket{\Psi(t)},
\end{equation}
where $H_0=\mathbf{p}^2/2-1/r$ is the field-free Hamiltonian and $V(t)$ the semi-classical interaction term. \cref{Eq: TDSE} is solved numerically, including the full electronic spectrum of the atom using the velocity gauge form for the interaction with the field, i.e. $V(t) = \hat p A(t)$. 
In addition, \cref{Eq: TDSE} is solved analytically using an essential-state approximation (two-level model). The laser field is linearly polarized along the z-direction, and it is defined by $E(t)=-\dot A$, with the potential given by $A(t)=A_0\Lambda_n(t) \sin(\omega_0 t + \phi)$, where $A_0$, $\Lambda_n(t)$, $\omega_0$ and $\phi$ are the amplitude, pulse envelope, central frequency and carrier-envelope phase (CEP), respectively. 
The time-dependent gauge-invariant exchange energy is given by the following expression, previously derived in Ref. \cite{stenquist_gauge-invariant_2023},
\begin{equation}
\label{Eq: DEt}
    \TAS_z(t)=-\int_{-\infty}^{t}dt'\,E(t')v_z(t'),
\end{equation}
where $t$ is time and $v_z(t')$ is the velocity expectation value. In the frequency domain, this yields the absorption 
\begin{equation}
\label{Eq: DEw}
\begin{split}
    \FTTAS_z(\omega) &= -2 \omega \,\mathrm{Im}[\tilde p_z(\omega)\tilde{A}^*(\omega)],
\end{split}
\end{equation}
where $\omega$ is the angular frequency, and $\tilde p_z(\omega)$ and $\tilde{A}(\omega)$ are the Fourier domain momentum expectation value and potential of the field, respectively. In \cref{fig:Absorption component} we present the Fourier transform of the field and the momentum expectation value. Here we observe how the overlap of these frequency-resolved quantities yields the Mollow-like absorption spectrum. The momentum expectation value is computed as 
\begin{equation}\label{Eq: pt general}
\begin{split}
    p_z(t) = \bra{\Psi(t)}\hat p_z \ket{\Psi(t)}. 
    \end{split}
\end{equation}
As described in Ref. \cite{wu_theory_2016} we introduce a filter function to limit the momentum expectation value in time, since Parseval's theorem requires that the Fourier transformed function is square integrable. Here, we use the pulse envelope as the filter function.

\begin{figure}[t!]
    \includegraphics[width = 0.45\textwidth]{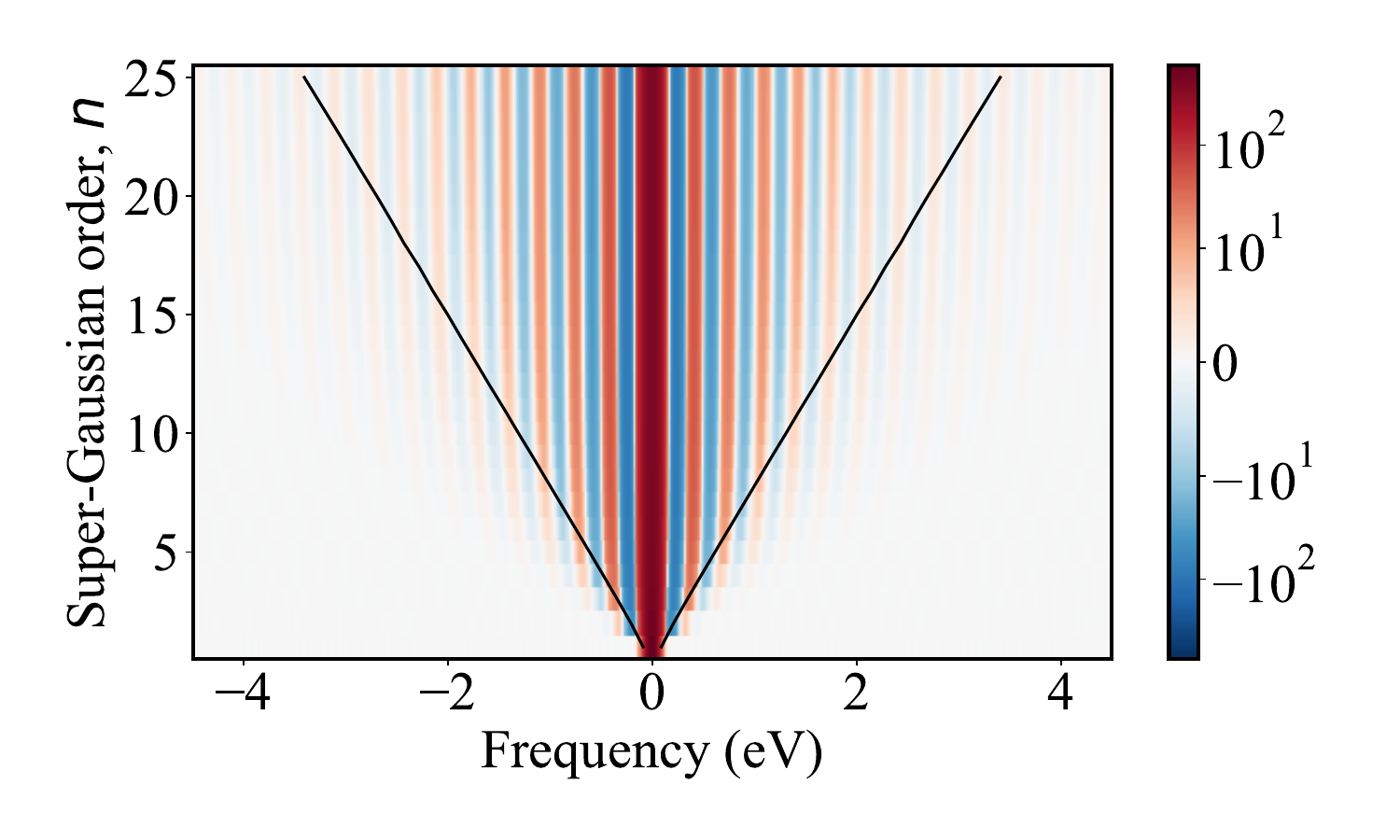}
\caption{Fourier transformed envelope, $\Lambda_n(\textcolor{red}{\omega})$, resolved over frequency and super-Gaussian order, $n$, with black lines denoting the energy width, $2\pi/\tau'_n$.}
\label{fig:Envelope width}
\end{figure}

We use a super-Gaussian sequence of pulses to quantify the energy exchange in Rabi cycling atoms, defined by the pulse envelope
\begin{equation}\label{Eq: envelope}
    \Lambda_n(t) = \exp \left[ -\frac{\text{ln}(2)}{2} \left(\frac{2t}{\tau}\right)^{2n}\right],
\end{equation}
where $\tau$ is the pulse duration expressed in full width at half maximum and $n$ is the super-Gaussian order. In the limit $n\rightarrow \infty$ the envelope becomes a flat top pulse
\begin{equation}\label{Eq: envelope ftop}
    \lim_{n\rightarrow\infty} \Lambda_n(t) = \theta(t-t_i) - \theta(t-t_f),
\end{equation}
where $\theta$ denotes the Heaviside functions, $t_i$ and $t_f$ are the initial and final time of the pulse. The centre of the pulse is defined at $t=0$. We define a measure of the Fourier width of the pulse, with modulations, as $2\pi/\tau'_n$, where $\tau'_n$ is the full-width half max of the derivative of the pulse (see Fig. 2 of the main article). This measure denotes where the Fourier transforms of the envelope is a few percent of its maximum value, as seen in \cref{fig:Envelope width}, where the Fourier transform of the envelope \cref{Eq: envelope} is resolved over frequency and the super-Gaussian order, $n$, where the black lines denote the energy spread defined through $\tau'_n$.

We have observed a minor CEP dependence in the case of the pulses with turn-on durations which are shorter than the Gaussian pulse due to the non-adiabatic initialization of the pulse. This is the most significant for the flat top pulse, where the turn-on time is instantaneous, yielding a maximum difference of a few percent of the total absorption. In the following, we therefore set the CEP to $\phi = 0$.

\subsection{Essential-state approximation}

The response to the XUV-FEL pulse can be calculated by applying an essential-state approximation which reduces the spectrum of the atom to a two-state problem. The wave function of this two-level model is then given by
\begin{equation}\label{Eq: state}
    \ket{\Psi(t)} = a(t) \ket{a} e^{-\ci \epsilon_a t} + b(t)\ket{b} e^{-\ci \epsilon_b t}, 
\end{equation}
where $a(t)$ and $b(t)$ are the time-dependent complex amplitudes of states $\ket{a}$ and $\ket{b}$, respectively, and $\epsilon_a$ and $\epsilon_b$ are their corresponding energies.
The time-dependent momentum expectation value in \cref{Eq: pt general} reduces to the expression, 
\begin{equation}\label{Eq: pt 2lvl}
\begin{split}
    p_z(t) = a^*(t)b(t)p_{ba}^*e^{-i\epsilon_{ba}t} + \text{c.c.}
    \end{split}
\end{equation}
where $p_{ba} = \bra{b}\hat p \ket{a} = i \epsilon_{ba} z_{ba}$ is the matrix element of the momentum operator, where  $\epsilon_{ba} = \epsilon_b - \epsilon_a$ is the resonance energy and $z_{ba} = \bra{b} \hat z \ket{a}$ is the transition dipole element. We consider the $1s-2p$ transition in hydrogen, which has a resonant energy of $\omega=10.2$\,eV, and transition dipole element $z_0=0.745$.

\subsubsection{Numerical solution} \label{Sec: 2lvl numerical}

The two-level problem can be solved numerically for an arbitrary pulse by inserting \cref{Eq: state} into \cref{Eq: TDSE} and expressing the time derivative of the wave function as $\frac{d}{dt}\Psi(t) = \frac{\Psi(t+dt) - \Psi(t-dt)}{2 dt}$. The state amplitudes can then be computed as 
\begin{equation}
    \begin{split}
        a(t+dt) =& \frac{2dt}{\ci} \left(\epsilon_a a(t) - \dot A(t) z_{ba} b(t) e^{-\ci \epsilon_{ba} t}\right) e^{i\epsilon_a dt} 
        \\ &+ a(t-dt)e^{\ci 2 \epsilon_a dt}, 
        \\
        b(t+dt) =& \frac{2dt}{\ci} \left(\epsilon_b b(t) - \dot A(t) z_{ba} a(t) e^{ \ci \epsilon_{ba} t}\right) e^{i\epsilon_b dt} 
        \\ &+ b(t-dt)e^{\ci 2 \epsilon_b dt}.
    \end{split}
\end{equation}

\subsubsection{Analytical solution with Hermitian Hamiltonian}

An analytical solution can be obtained in the case of a flat top envelope, \cref{Eq: envelope ftop}.  Inserting \cref{Eq: state} and the potential corresponding to a flat top XUV-FEL pulse into \cref{Eq: TDSE}, within the rotating wave approximation (RWA), one obtains the well-known Rabi oscillation amplitudes,
\begin{equation}\label{Eq: Rabi amp}
    \begin{split}
        a( t) &= e^{ \frac{\ci \Det  t}{2}} \left[ a_0 \cos(\frac{W t}{2}) - \ci \left(\frac{\Det a_0 + \ci \Omega b_0e^{\ci\phi }}{W}\right)\sin(\frac{W t}{2}) \right], \\
        b( t) &= e^{-\frac{\ci \Det  t}{2}} \left[ b_0 \cos(\frac{W t}{2}) + \ci \left(\frac{\Det b_0 + \ci \Omega a_0e^{-\ci\phi}}{W}\right)\sin(\frac{W t}{2}) \right],
    \end{split}
\end{equation}
where $a_0$ and $b_0$ are the inital values of states $\ket{a}$ and $\ket{b}$, respectively, $\Omega = z E_0$ is the Rabi frequency, $\Det$ is the detuning, defined by $\omega_0 = \epsilon_{ba} + \Det$, where $\omega_0$ is the central frequency of the field. The generalized Rabi frequency is $W = \sqrt{\Omega^2 + \Det^2}$. 
Inserting, \cref{Eq: Rabi amp} into \cref{Eq: pt 2lvl} the time-dependent momentum expectation value is 
\begin{equation}\label{Eq: mom ExVal t ana}
\begin{split}
p_z(t) = & p_{ba}^*\big[ A_- + A_+ \cos(W(t-t_i)) \\
         & + \ci B\sin(W(t-t_i)) \big] e^{-\ci (\epsilon_{ba} + \Det) t} + \text{c.c.}, 
\end{split}
\end{equation}
where coefficients are given by 
\begin{equation}\label{Eq: p_t coeff}
\begin{split}
    A_\pm =& \frac{1}{2} \Big[ a_0^*b_0 \pm \frac{1}{W^2} \big(\Det^2a_0^*b_0 - \Det \Omega\abs{a_0}^2e^{-\ci\phi} \\
           & + \Det \Omega^*\abs{b_0}^2e^{-\ci\phi} - \Omega^2 a_0^*b_0 e^{-\ci2\phi} \big) \Big], \\
    B     =& \frac{1}{2W} (2 \Det a_0^*b_0 - \Omega\abs{a_0}^2e^{-\ci\phi} + \Omega^*\abs{b_0}^2e^{-\ci\phi}  ).
\end{split}
\end{equation}

The Fourier transform of the 
flat top potential is
\begin{equation} \label{Eq: vec pot w}
    A(\omega) = \frac{\ci A_0 }{\sqrt{2\pi}} \frac{\tau}{2} \sinc\left((\omega-\epsilon_{ba}-\Det)\frac{\tau}{2}\right),
\end{equation}
and the momentum expectation becomes
\begin{equation} \label{Eq: p_w}
\begin{split}
    p(\omega) 
    &= \frac{p_{ba}^*\tau}{\sqrt{2\pi}} 
    \Bigg [
    A_-\, \, \sinc
    \left((\omega-\epsilon_{ba}-\Det)\frac{\tau}{2}\right) \\
    &+ \frac{(A_+ + B)}{2} \sinc\left((\omega+W-\epsilon_{ba}-\Det)\frac{\tau}{2}\right) e^{-\ci W t_i} \\
    &+ \frac{(A_+ - B)}{2} \sinc\left((\omega-W-\epsilon_{ba}-\Det)\frac{\tau}{2}\right) e^{\ci W t_i} 
    \Bigg ],
\end{split}
\end{equation}
where for both \cref{Eq: vec pot w} and \cref{Eq: p_w} we have neglected the negative frequency components. In the case of initial state $\abs{a_0}^2 = 1$, $\abs{b_0}^2 = 0$ and $\Det = 0$ the constant vanishes $A_\pm = 0$, yielding a two peak structure in the momentum spectrum, as seen in \cref{fig:Absorption component}, with the central frequency component of the Mollow-like absorption triplet being associated to the main peak of the Fourier transform of the field. 

\textcolor{white}{blank}

\begin{widetext}

Inserting \cref{Eq: vec pot w} and \cref{Eq: p_w} into \cref{Eq: DEw} and expressing the pulse duration $\tau = NT = 2 \pi N / W $ in terms of the number of Rabi periods $N$, the frequency domain absorption becomes 
\begin{equation}\label{DE w}
    \begin{split}
        \FTTAS_z(\omega) = \frac{\omega }{2 \pi } A_0 \tau^2 \epsilon_{ba} z_{ba} \text{Im}\Bigg[ &A_- \sinc^2\left(\frac{(\omega-\epsilon_{ba}-\Delta\omega)N \pi}{W}\right)
\\ + 
&\frac{(A_+ + B)}{2}\sinc\left(\frac{(\omega+W-\epsilon_{ba}-\Delta\omega)N \pi}{W}\right)
\sinc\left(\frac{(\omega-\epsilon_{ba}-\Delta\omega)N \pi}{W}\right)e^{-\ci W t_i} 
\\ + 
&\frac{(A_+ - B)}{2}\sinc\left(\frac{(\omega-W-\epsilon_{ba}-\Delta\omega)N \pi}{W}\right)
\sinc\left(\frac{(\omega-\epsilon_{ba}-\Delta\omega)N \pi}{W}\right)e^{\ci W t_i} 
\Bigg],
    \end{split}
\end{equation}
note that $z_{ba} \in \mathbb{R}$. By limiting the initial state to $\abs{a_0}^2 = 1$, $\abs{b_0}^2 = 0$, we can further simplify this expression onto the form $\FTTAS(\omega) = \FTTAS^+(\omega) + \FTTAS^-(\omega)$,
\begin{equation}\label{DE w simple}
    \begin{split}
&\FTTAS^\pm(\omega) = \frac{\pi \epsilon_{ba} z_{ba} \omega A_0 \Omega N^2 }{2 W^3}  \sin(W t_0) 
\left(1\pm\frac{\Delta\omega}{W}\right) 
\sinc\left[\frac{(\omega-\epsilon_{ba}-\Det)N \pi}{W}\right]
\sinc\left[\frac{(\omega-\epsilon_{ba}\pm W-\Det)N \pi}{W}\right].
    \end{split}
\end{equation}
The term $\FTTAS^+(\omega)$ yielding two peaks centred on $\{\epsilon_{ba} - W-\Det, \,\epsilon_{ba}-\Det\}$, and $\FTTAS^+(\omega)$ two peaks centred on $\{ \epsilon_{ba}-\Det, \, \epsilon_{ba} + W-\Det\}$.
\end{widetext}

Together they form the characteristic Mollow triplet. Detuning yields an energy shift of the peaks and the asymmetry in peak amplitude due to the factor $(1\pm\Det / W)$. On resonance, the sum of the two terms in \cref{DE w simple} will yield the amplitude ratio between the central and the side peaks of 2:1. Initially occupying the excited state gives the opposite behaviour of initially occupying the ground state. It is not possible to choose the initial state in such a way that you only get two out of the four components. 

\subsubsection{Analytical solution with non-Hermitian Hamiltonian} 

By using an effective Hamiltonian formulation \cite{cohentannoudji_atomphoton_1998} we can include effects beyond the essential-state approximation, using scaling parameters from Ref.~\cite{olofsson_photoelectron_2023}. In this manner, a quantum optics framework is used to solve the TDSE for two states using an effective non-Hermitian Hamiltonian given by
\begin{equation}
H_\text{eff} = 
    \begin{bmatrix}
        h_{aa}&h_{ab}\\
        h_{ba}&h_{bb}
    \end{bmatrix},
\end{equation}
where the matrix elements are $h_{aa} = E_a + R_{aa}$, $h_{bb} = E_b + R_{bb}$ and $h_{ab} = h_{ba} = R_{ba}$. $E_a$ and $E_b$ being the dressed state energies of states $\ket{a}$ and $\ket{b}$, respectively, and $R_{aa}$, $R_{bb}$, and $R_{ba}$ are the corrections detailed in Ref.~\cite{olofsson_photoelectron_2023}. 

The TDSE with the effective Hamiltonian can be solved numerically, as described above, or analytically by determining the eigenvalues
\begin{equation}
\lambda_\pm = \frac{h_{aa}+h_{bb}}{2} \pm \frac{1}{2} \sqrt{(h_{aa}-h_{aa})^2 + 4 h_{ab}^2}    ,
\end{equation}
and eigenstates 
\begin{equation}
    \ket{\pm} = h_{ab} \ket{a} + \left[\frac{h_{bb}-h_{aa}}{2} \pm \frac{1}{2} \sqrt{(h_{aa}-h_{aa})^2 + 4 h_{ab}^2}\right] \ket{b},
\end{equation}
of the effective Hamiltonian. 
The complex generalized Rabi frequency is described as 
\begin{equation}
    W = \sqrt{(h_{aa}-h_{bb})^2+4h_{ab}^2}.
\end{equation}
The wave function of the two level system is then given by 
\begin{equation}
\ket{\Psi} = c_+\ket{+}e^{-i\lambda_+ t} + c_-\ket{-}e^{-i\lambda_- t}
\end{equation}
where the coefficients $c_+,c_-$ are time-independent coefficients given by the initial state of the atom
\begin{equation}
    \begin{split}
a(t=0) &= \bra{a}\ket{\Psi(t=0)}, \\
b(t=0) &= \bra{b}\ket{\Psi(t=0)}.
    \end{split}
\end{equation}

The time-dependent momentum expectation value is then calculated correspondingly to \cref{Eq: pt 2lvl}, yielding 
\begin{equation}
\label{Eq:Heff_pt}
    p(t) = \sum_j^4 A_j p_{ba}^*e^{-B_j t_0} e^{B_j t -i (\epsilon_{ba}+\Det) t}  
\end{equation}
where $A_{s\pm} = C_\pm^*D_\pm$ and $A_{c\pm} = C_\pm^*D_\mp$  
are constants depending on the initial state of the system, where $C_\pm = c_\pm h_{ba}$, $D_\pm = c_\pm [(h_{aa}-h_{bb})/2 \pm W / 2]$, and $B_j$ is defined as  
\begin{equation}
    \begin{split}
B_{s\pm} &= \pm i\text{Re}(W) + \text{Im}(h_{aa} + h_{bb}) , \\
B_{c\pm} &= \pm \text{Im}(W) + \text{Im}(h_{aa} + h_{bb}), 
    \end{split}
\end{equation}
where the subscripts $s$ and $c$ denote the sideband and central peak, respectively. 
The momentum expectation value in the frequency domain is
\begin{equation}
    p(\omega) = 
    \sum_j^4 \frac{A_j}{\sqrt{2\pi}} p_{ba}^*e^{-B_j t_0} \tau \,\sinc\left[(\omega-\epsilon_{ba}- \Det -iB_j)\frac{\tau}{2}\right].
\end{equation}
which is reminiscent of the Hermitian-Hamiltonian case \cref{Eq: p_w}, however in this case we have four terms
expressed as sinc functions with complex arguments.

In the long pulse limit $\tau\rightarrow \infty$ we get the momentum expectation value
\begin{equation}
    p(\omega) = \sum_j^4 i \frac{A_j \,p_{ba}^*}{\sqrt{2\pi}} \frac{1}{\omega- \epsilon_{ba} - iB_j},
\end{equation}
which is Lorentzian, in agreement with the previous studies on resonant fluorescence \cite{mollow_power_1969}, with the width of the peaks being determined by the lifetimes of the states. The imaginary part of the momentum is symmetrical and the real part is asymmetrical. 

\subsection{Comparison of the models}
\begin{figure}[t!]
    \includegraphics[width = 0.49\textwidth]{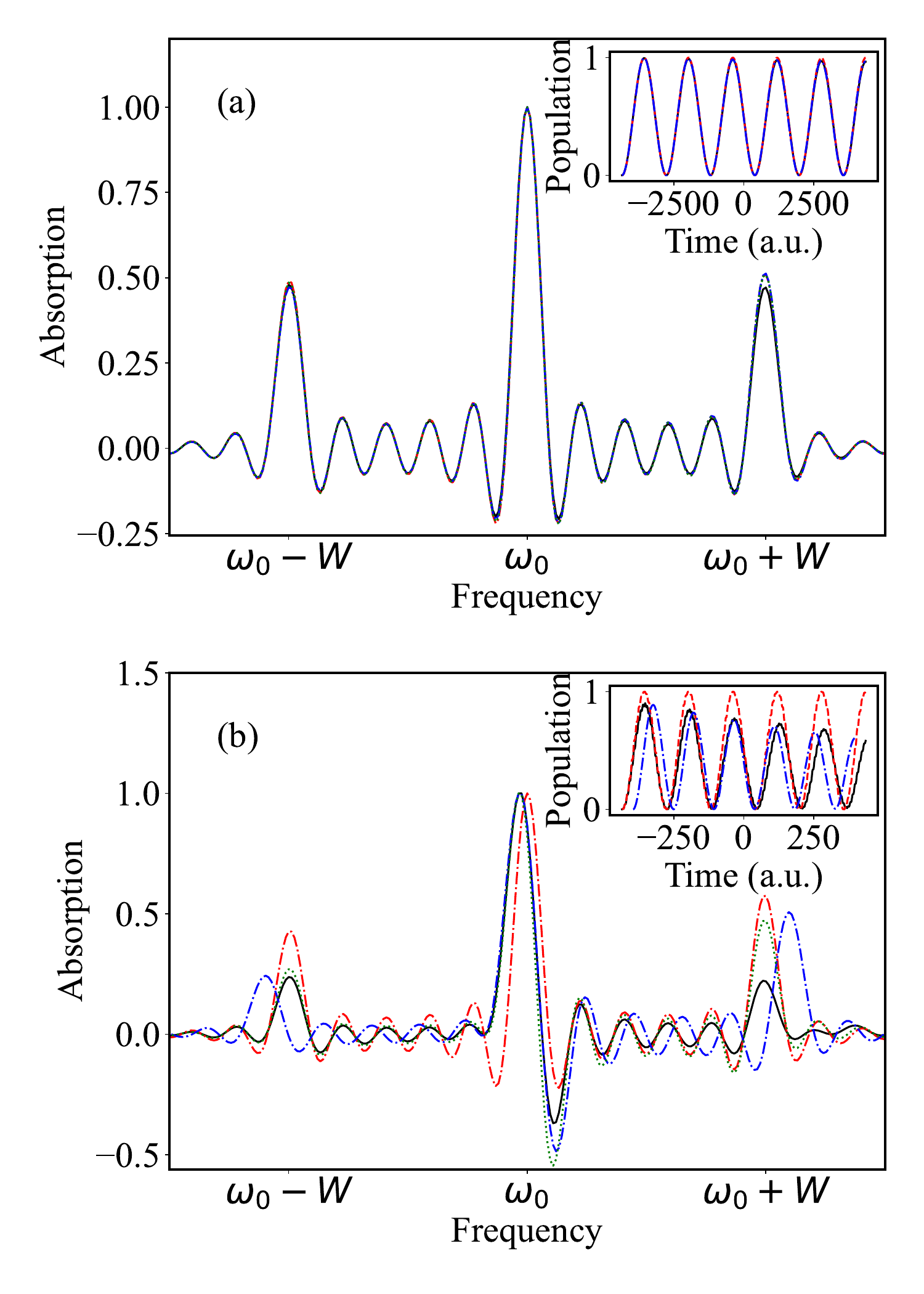}
\caption{Comparison between the different models for the absorption at low intensity $I=10^{12}$ Wcm$^{-2}$ (a), and high intensity $I=10^{14}$ Wcm$^{-2}$ (b). Exact calculations (black), two-level essential state model (red dashed), effective Hamiltonian model (blue dash-dotted) and reduced exact calculations (green dotted) are compared. The inlays show the excited state population over time of the corresponding methods.}
\label{fig:method compare}
\end{figure}

The comparison of the introduced models is presented in \cref{fig:method compare}, for low intensity $I=10^{12}$ Wcm$^{-2}$ (a) and high intensity $I=10^{14}$ Wcm$^{-2}$ (b). The exact numerical calculation is shown in black lines. In the low-intensity regime (a), the Hermitian model (red dashed) accurately reconstructs the absorption spectra, being accurate below the intensity $I = 10^{13}$ Wcm$^{-2}$. The excited state population of the corresponding methods is presented in the insert, where good agreement is observed.  
\begin{figure*}[t!]
    \includegraphics[width = 0.99\textwidth]{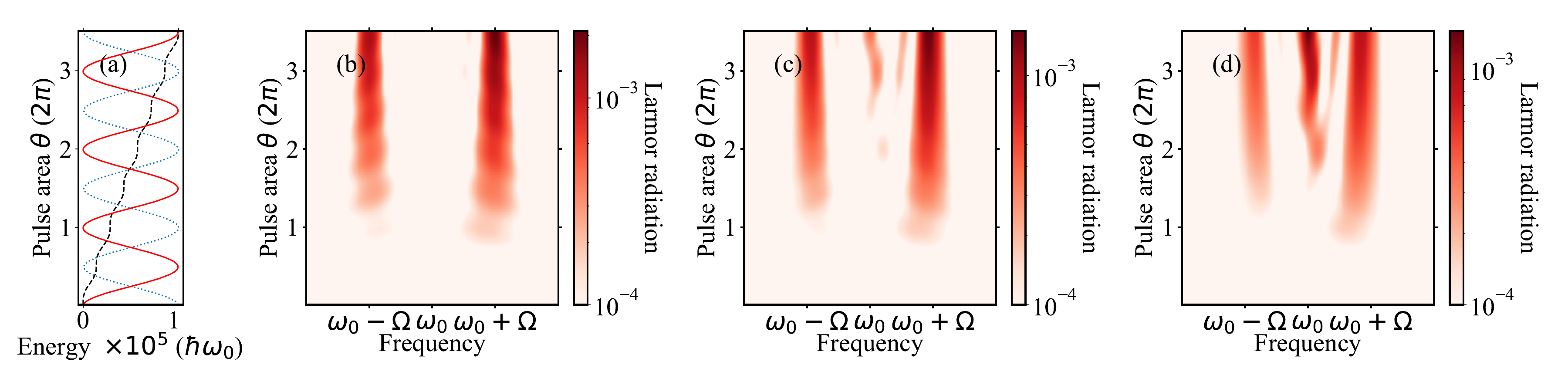}
\caption{Build-up of the Larmor radiation spectrum. The frequency and angle-integrated Larmor radiation for a flat top pulse, is presented in (a) in black dashed lines, along with the population of the excited state in full red lines. The frequency-resolved, angle-integrated Larmor radiation spectra are presented in (b), (c) and (d), for a flat-top, super-Gaussian and Gaussian envelope, respectively.}
\label{fig:Larmor buildup}
\end{figure*}

However, for high intensities the non-essential states of the electronic spectrum become significant, necessitating the use of the effective Hamiltonian model (blue dash dotted). This model reconstructs the Fano-like shape of the central peak of the absorption spectra at high intensities (b). 
There is a discrepancy in the effective model, having populations which oscillate faster, see insert, due to the 3p state being accessible from the 1s state \cite{paulisch_beyond_2014}. This yields a higher Rabi frequency, and hence a larger separation between the absorption peaks. This can be resolved by including higher order corrections in the energy ($z$) of the level-shift operator, as proposed within the resolvent formalism in chapter 7.5.2 of Ref. \cite{faisal_theory_1987}. 
To compare, we match the number of Rabi cycles, $\theta / 2\pi = 5.5$, that the atom undergoes for all our methods. 

Since the Fano-like shape of the absorption at high intensity is recreated by the effective Hamiltonian essential-state model, we deduce that it is due to interaction with the other states, but not due to the other states themselves. This is further ascertained by projecting the solution of the TDSE on the essential states  $\ket{\widetilde\Psi(t)} = (\ket{a}\bra{a} + \ket{b}\bra{b}) \ket{\Psi(t)}$ and calculating the momentum expectation value with the reduced wave function $p_z^\text{red}(t) = \bra{\widetilde \Psi(t)}\hat p_z \ket{\widetilde \Psi(t)}$. The absorption computed using the reduced wave function is presented in green dotted lines. Interestingly, the reduced absorption, the effective Hamiltonian model and the Hermitian two-state model all show asymmetry in the sidebands, which disappears when considering the full electronic spectra.



\section{Emission Formalism}

\subsection{Larmor radiation}

The Larmor radiation is classically described as the light emitted from an accelerating charge. Resolved over angle and energy it is given by
\begin{equation}
    W(\omega;\theta) = \frac{\sin^2(\theta)\abs{\tilde a(\omega)}^2}{2\pi c^3}, 
\end{equation}
the total, angle-integrated, energy-resolved radiation is given by 
\begin{equation} \label{Eq: full Larmor}
    W(\omega) = \frac{4\abs{\tilde a(\omega)}^2}{3 c^3},
\end{equation}
where the acceleration expectation value is given by $a(t) = \dot v(t)$.

The buildup of the Larmor radiation spectra is presented in \cref{fig:Larmor buildup}, where in (a) we see the linear-like increase of the emited energy (black dashed lines). 
In contrast, to the absorption spectra, the frequency-resolved Larmor radiation does not show a dependence on the final state. For interaction with a flat top pulse in (b) we observe two separate and steady bands being formed as the area of the pulse is increased. The position of the Larmor bands agrees with the position of the Mollow-triplet sidebands.  
In the super-Gaussian (c) and Gaussian cases (d) the Larmor radiation also develops a weaker main peak with some interference structure visible in the spectra for areas beyond 2 Rabi cycles. 
%

\subsection{Resonance fluorescence}

We follow the resonant fluorescence formalism presented in \cite{florjaczyk_resonance_1985}, built on a two-state model described in a quantum optics framework. The fluorescence spectrum is composed of two parts, $S(\omega,\Gamma_F) = S_s + S_q$, referred to as the "semiclassical" or "coherent" part $S_s$ and the "quantum" or "incoherent" part $S_q$. These are defined as 
\begin{equation}
    \begin{split}
        S_s(\omega,\Gamma_f) &= 2 \Gamma_F \abs{ \int dt e^{[\Gamma_F - i(\omega-\omega_0)]t} a(t)b^*(t) }^2, \\
        S_q(\omega,\Gamma_f) &= 2 \Gamma_F \abs{ \int dt e^{[\Gamma_F - i(\omega-\omega_0)]t} \abs{b(t)}^2 }^2,
    \end{split}
\end{equation}
where $\Gamma_f$ is a filter parameter giving the spectral resolution of the spectrometer, and $\Gamma_f^{-1}$ is the accumulation time of the spectrometer. The filter parameter is chosen to be small, yielding a well resolved fluorescence spectrum. This formulation accurately reproduces the shape of the fluorescence spectrum \cite{boos_signatures_2024}, but the magnitude of the signal is arbitrary. 

\begin{figure}
    \includegraphics[width = 0.48\textwidth]{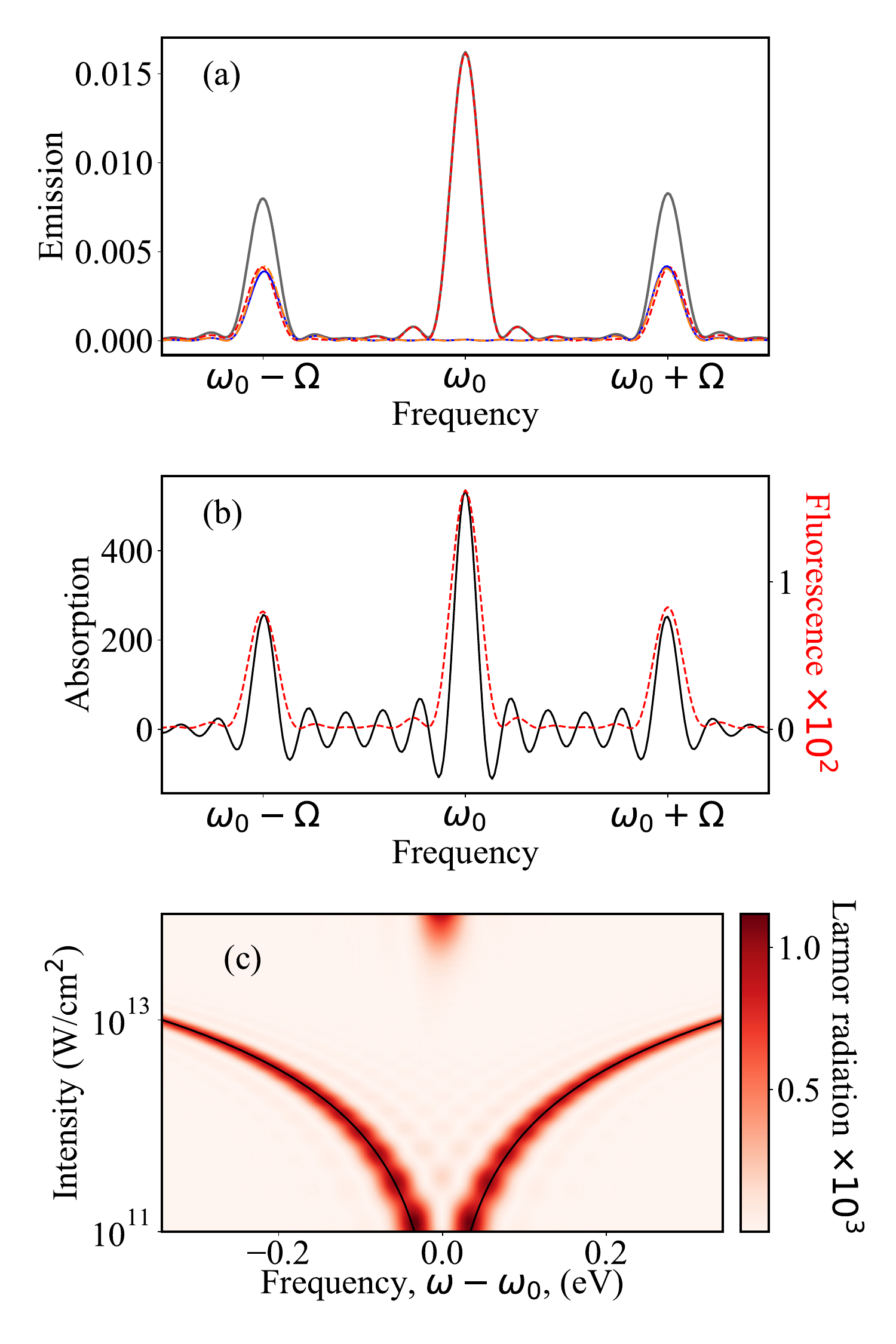}
\caption{Comparison of the Fluorescence spectra (grey), its semiclassical (red dashed) and quantum (blue) components with the Larmor radiation (orange dash-dotted) (a). The fluorescence is scaled to the Larmor radiation and is compared to the absorption in (b). The Larmor radiation, resolved over frequency and intensity, is presented in (c).}
\label{fig:emission results}
\end{figure}

As the resonance fluorescence formalism is rooted in the essential state approach, it is only valid in the low-intensity regime, where its applicability has been verified. The two emission approaches are compared in \cref{fig:emission results}(a) where the fluorescence spectrum is shown in grey and its components: the quantum part and semiclassical part, are shown in red dashed and blue full lines, respectively. The Larmor radiation, integrated over solid angle, see \cref{Eq: full Larmor}, is presented in dash-dotted orange lines and is observed to match the shape of the semiclassical part of the fluorescence. This is expected, as the Larmor radiation is the emission from a classically accelerated charge. As the magnitude of the Larmor radiation is physically meaningful, we scale the fluorescence to match it, note that this is done in Fig. 1 of the main article. In (b) we compare the absorption with the fluorescence, where we observe that the position and relative height of the peaks are in good agreement, the Fluorescence is seen to be much smaller than the absorption. As opposed to the Fluorescence, the Larmor radiation is not formulated in an essential-state model and is therefore valid for high intensity also. We present the intensity- and frequency-resolved Larmor radiation in (c).

\bibliography{Mollow_ref}